\documentstyle[12pt]{article}

\newcommand{\nc}{\newcommand}
\nc{\beq}{\begin{equation}}
\nc{\eeq}{\end{equation}}
\nc{\beqa}{\begin{eqnarray}}
\nc{\eeqa}{\end{eqnarray}}

\def\gsim{\mathrel{\rlap{\lower4pt\hbox{\hskip1pt$\sim$}}
    \raise1pt\hbox{$>$}}}       

\input epsf
\newwrite\ffile\global\newcount\figno \global\figno=1

\def\writedef#1{}
\def\figin{\epsfcheck\figin}\def\figins{\epsfcheck\figins}
\def\epsfcheck{\ifx\epsfbox\UnDeFiNeD
\message{(NO epsf.tex, FIGURES WILL BE IGNORED)}
\gdef\figin##1{\vskip2in}\gdef\figins##1{\hskip.5in}
\else\message{(FIGURES WILL BE INCLUDED)}%
\gdef\figin##1{##1}\gdef\figins##1{##1}\fi}
\def\figinsert{}
\def\ifig#1#2#3{\xdef#1{fig.~\the\figno}
\writedef{#1\leftbracket fig.\noexpand~\the\figno}%
\figinsert\figin{\centerline{#3}}\medskip\centerline{\vbox{\baselineskip12pt
\advance\hsize by -1truein\center\footnotesize{  Fig.~\the\figno.} #2}}
\bigskip\endinsert\global\advance\figno by1}
\def\endinsert{}

\begin{document}



\title{\large{\bf Entropy Bounds and Dark Energy}}

\author{Stephen D.H.~Hsu\thanks{Permanent address: Institute of
Theoretical Science and Department of Physics, University of
Oregon, Eugene OR 97403.
Email: hsu@duende.uoregon.edu} \\
\\
Department of Physics \\
California Institute of Technology, Pasadena, CA 91125\\
}



\maketitle

\begin{picture}(0,0)(0,0)
\end{picture}
\vspace{-24pt}

\begin{abstract}
Entropy bounds render quantum corrections to the cosmological
constant $\Lambda$ finite. Under certain assumptions, the natural
value of $\Lambda$ is of order the observed dark energy density
$\sim 10^{-10} \, {\rm eV}^4$, thereby resolving the cosmological
constant problem. We note that the dark energy equation of state
in these scenarios is $w \equiv p / \rho = 0$ over cosmological
distances, and is strongly disfavored by observational data.
Alternatively, $\Lambda$ in these scenarios might account for the
diffuse dark matter component of the cosmological energy density.
\end{abstract}


\newpage


There is evidence that gravity limits the number of quantum states
accessible to a system, yielding non-extensive or {\it
holographic} entropy bounds \cite{Bekenstein,Thooft,Susskind,FSB}.
These bounds require that the entropy of a region increase less
rapidly than its volume, generally as its area in Planck units.
Hence, they imply that the dimension of the Hilbert space (number
of degrees of freedom) describing a region is finite and much
smaller than previously expected.

Entropy bounds have immediate implications for the cosmological
constant problem \cite{ccp}. In conventional quantum field theory
(in which the entropy is extensive), the quantum corrections to
the vacuum energy are typically divergent. An extreme fine-tuning
of bare parameters, for which no satisfactory mechanism is known,
is required to keep the vacuum energy within observed limits.

As noted by several authors \cite{CKN,Thomas}, there is a
connection between the entropy of a system and the quantum
correction to the vacuum energy. In $d=4$ field theory, the
classical vacuum energy density receives quantum corrections \beq
\Lambda = \Lambda_0 + {\cal O} (s^{4/3}) ~~,\eeq where $s = S/V$
is the entropy density per unit volume. This is because both
$\Lambda$ and $s$ are dominated by ultraviolet modes; indeed, in
the simplest calculation the vacuum energy correction is simply
the zero point energy summed over all modes: \beq \label{Lqm}
\Lambda_{qm} ~\sim ~ \int^{M} d^3k \, \sqrt{\vec{k}^2 + m^2} ~\sim
~ M^4~~,\eeq where $M$ is the UV cutoff. The corresponding entropy
density is $s \sim M^3$. Rendering the entropy density finite also
renders the correction to the cosmological constant finite.

The naive estimate in (\ref{Lqm}), often used to characterize the
severity of the cosmological constant problem, is likely to be
modified by gravitational effects when we consider length scales
of relevance to cosmology. The highest energy states of a system
allowed by the cutoff $M$: $E \sim M^4 L^3$ for a system of size
$L$, are already within their Schwarzschild radius if $L < R_s
\sim E$, or $L > M^{-2}$. (We use Planckian units, where Newton's
constant is unity.) One can see the self-gravitational effects of
the vacuum energy explicitly in perturbation theory as follows.
Diagrammatically, the usual contribution to $\Lambda_{qm}$ in
(\ref{Lqm}) is given by a vacuum bubble. Treating the graviton in
perturbation theory, there is a correction to the vacuum energy
from the connected (but not 1PI) graph with a graviton line
connecting two bubbles. This graph is most easily evaluated in
coordinate space, and has the form $M^8 L^2$. It is a large
correction to the single vacuum bubble when $M^4 L^2 \sim 1$.
Additional graphs containing $g$ graviton lines and $g+1$ bubbles
contribute $M^4 (M^4 L^2)^g$ to the vacuum energy. They show that
when $L > M^{-2}$ there is a large gravitational back reaction. To
eliminate these graphs one needs to shift the metric, presumably
about a classical de Sitter background.

One of the main ideas leading to holography is that black hole
states must be treated more carefully in quantum gravity. A
correct evaluation of $\Lambda_{qm}$ could yield a result which is
much smaller than (\ref{Lqm}), and dependent on length scale $L$ .
This effect is clearly related to the entropy bounds resulting
from gravity. By making specific assumptions, one can estimate the
natural size of the correction $\Lambda_{qm}$.

In \cite{CKN}, a relationship between the size $L$ of the region
under consideration (which provides an IR cutoff) and the UV
cutoff $M$ was assumed. The relationship is deduced by requiring
that no state in the Hilbert space have energy $E$ so large that
the Schwarzschild radius $R_s \sim E$ exceeds $L$. Under this
requirement, the entropy grows no faster than $A^{3/4} \sim
L^{3/2}$ \cite{Thooft}, where $A$ is the area of the region. In
physical terms, this corresponds to the assumption that the
effective field theory ${\cal L} (L,M)$ describe all states of the
system {\it excluding} those for which it has already collapsed to
a black hole. Further, it is assumed that the black hole states do
not contribute to $\Lambda_{qm}$. Under these assumptions we
obtain \beq \label{LCKN} \Lambda_{qm} ~\sim~ s^{4/3} ~\sim~ \Big(
{L^{3/2} \over L^3} \Big)^{4/3} ~\sim~ L^{-2} ~~.\eeq  Note, the
value of $M$ obtained below satisfies $m_i > M$ for all standard
model particles $i$ except the photon and perhaps the neutrinos.
For these particles the result of (\ref{Lqm}) is modified to
$\Lambda_{qm} \sim \sum_i \, m_i \, M^3$, and the corresponding
relationship between $L$ and $M$ more complicated than described
above. Nevertheless the relationship between $\Lambda_{qm}$ and
$L$, which is central to what follows, remains the same.

In \cite{Thomas}, it was assumed that the entropy bound has the
usual area form: $S < A$, but that the delocalized states of the
system have typical Heisenberg energy $\sim 1/L$. This yields \beq
\label{LT} \Lambda_{qm} ~\sim~ {s \over L} ~\sim~ { L^2 \over L^3
L} ~\sim~ L^{-2} ~~,\eeq which is the same scaling as in
\cite{CKN}, but based on different assumptions. Evaluating
(\ref{LCKN}),(\ref{LT}) using the size of the observed universe
(the current horizon size $L_{\rm today} \sim 10 \,{\rm Gy}$)
yields a result $\Lambda_{qm} \sim 10^{-10} \, {\rm eV}^4$, which
would explain the observed dark energy density \cite{WMAP},
assuming that $\Lambda_0 \sim \Lambda_{qm}$. (Note, using the area
law $S < A$ to determine the $L,M$ relation in \cite{CKN} yields a
much larger estimate of $\Lambda_{qm}$. However it seems quite
plausible that the black hole states excluded in the $A^{3/4}$
entropy bound do not contribute to the vacuum energy in the usual
way.)

While the holographic ideas discussed above yield the correct
value of the observed cosmological constant, they do not yield the
correct equation of state. Consider the universe at some earlier
time when the horizon size was $L$ ($L < L_{\rm today}$). By
causality, gravitational influences have not had time to propagate
between regions separated by more than $L$. Therefore, the vacuum
energy which appears in the Einstein equations, driving the
instantaneous expansion, is $\Lambda (L)$.  However, because the
cosmological constant is $L$-dependent, the dark energy equation
of state $w \equiv p / \rho$ is not equal to $-1$. During the
matter dominated epoch to which the WMAP and supernova
measurements are sensitive, the horizon size grows as the
Robertson-Walker scale factor $R(t)^{3/2} \,$, so (\ref{LCKN}) and
(\ref{LT}) imply \beq \label{LR} \Lambda(L) ~\sim~ R(t)^{-3}~~,
\eeq or $w = 0$ at the largest scales (recall, for equation of
state $w$ the energy redshift behavior is $\rho(t) \sim
R(t)^{-3(1+w)}$). The WMAP data, which are sensitive to $\Lambda$
over a redshift range of roughly $10^3$ (since decoupling), imply
$w < -0.78$ (95\% CL) \cite{WMAP}. In other words, the data
require a cosmological constant that is much more constant than
obtained in holographic scenarios. In fact, in the scenarios
\cite{CKN,Thomas} $\Lambda(L)$ is at all times comparable to the
radiation + matter energy density, which would be problematic for
structure formation \cite{Turner}. More generally, if we take the
entropy bound $S < c \, L^{n}$ and assume that the dark energy
$\Lambda(L) \sim s^{4/3}$, the data requires that $n > 2.7$ (95\%
CL). This does not rule out holography per se, nor a holographic
improvement to the fine tuning problem, but does rule out a simple
connection between dark energy and holography.

It is difficult to see how holographic ideas can avoid this
problem with the equation of state. By linking IR and UV scales
$L$ and $M$ through entropy bounds, holography does provide an
essential ingredient long believed necessary to solve the
cosmological constant problem\footnote{The solution proposed in
\cite{Horava}, which promotes the cosmological constant to a
dynamical field, implies that our universe's groundstate has zero
cosmological constant, but does not explain the observed dark
energy density. Banks \cite{Banks} and Fischler \cite{Willy} have
proposed that the cosmological constant is not a dynamical
consequence of quantum fluctuations, but rather an input parameter
related to the number of degrees of freedom in the universe.}.
However, observations indicate that the dark energy density is
varying quite slowly (if at all) with the size of our universe.

An alternative possibility is that $\Lambda (L)$ is actually the
diffuse dark matter, with $w = 0$, while dark energy has some
other origin. Dark matter is roughly 30 percent of the total
energy density of the universe - within a factor of two of the
dark energy density itself. It is an open question whether
$\Lambda (L)$ would behave on sub-horizon length scales as
ordinary dark matter (for example, clumping during structure
formation) or rather as a smooth component of energy density.

Finally, we describe another bound related to the possibility that
$\Lambda (L)$ might be a function of lengthscale\footnote{The
author thanks M. Kamionkowski and B. Murray for discussions on
this subject.}. Let \beq \label{local} \Lambda(L) \sim 10^{-10} \,
{\rm eV}^4 \left( {L_{\rm today} \over L}  \right)^k~~, \eeq where
$L_{\rm today}$ is the current horizon size. It is possible that
entropy bounds only restrict $\Lambda$ on cosmological length
scales, but there is no obvious reason that the holographic ideas
cannot be applied to smaller subsystems in our universe. For
smaller $L$ the corresponding cosmological constant is larger, and
leads to a repulsive force $F(L) \sim \Lambda(L) L$. Perhaps the
best limit on such a force comes from planetary motion in our
solar system, for which $L_{\rm solar~system} \sim 50~{\bf AU}$. A
conservative bound on $k$ can be deduced by requiring that $F(L)$
be much less than the the sun's gravitational pull: $F(L) <<
M_{\odot} / L^2$. We find that $k = 2$ is ruled out by more than
10 orders of magnitude, and that $k = 1$ is just allowed. Note
that if holographic effects are not responsible for the dark
energy, the overall coefficient in (\ref{local}) might be much
smaller and the corresponding bound on $k$ much weaker. It is also
possible that the manifestation of holographic ideas is more
subtle than described here \cite{Banks,Willy}.

\bigskip

\noindent The author would like to thank M. Graesser, M.
Kamionkowski, B. Murray and M. Wise for comments, and the Caltech
theory group for its hospitality. This work was supported in part
under DOE contract DE-FG06-85ER40224.


\bigskip


\end{document}